\documentclass[aps,twocolumn,showpacs,showkeys,nofootinbib,superscriptaddress]{revtex4-1}
\usepackage{graphicx}     
\usepackage{amsmath}
\usepackage{amssymb, amsfonts}
\usepackage{array}
\usepackage{braket}
\usepackage{cases}
\usepackage[utf8]{inputenc}
\usepackage[overload]{empheq}
\usepackage{relsize}  
\usepackage{multirow}
\usepackage{dcolumn}  
\usepackage{hyperref}
\usepackage{xcolor}
\usepackage{relsize}%
\hypersetup{
	colorlinks   = true, 
	urlcolor     = blue, 
	linkcolor    = blue, 
	citecolor    = blue 
}
\newcommand{\RN}[1]{%
	\textup{\uppercase\expandafter{\romannumeral#1}}%
}
\newcommand{\cev}[1]{\reflectbox{\ensuremath{\vec{\reflectbox{\ensuremath{#1}}}}}} 
\newcommand{\angstrom}{\text{\normalfont\AA}}
\newcolumntype{d}[1]{D{.}{.}{#1}}
\newcolumntype{K}[1]{>{\centering\arraybackslash}p{#1}}

\usepackage{floatflt}     
\usepackage{wrapfig}
\usepackage{color}
\usepackage{pstricks}

\makeatletter
\newcommand*{\rom}[1]{\expandafter\@slowromancap\romannumeral #1@}
\usepackage{color}
\definecolor{DarkRed}{rgb}{0.35,0.01,0.01}
\definecolor{Linen}{rgb}{0.98,0.98,0.94}
\definecolor{Blue}{rgb}{0.,0.,1.0}
\definecolor{DarkBlue}{rgb}{0.099,0.099,0.44}
\definecolor{DarkGreen}{rgb}{0.0,0.4,0.0}
\definecolor{Turquoise}{rgb}{0.0,0.9,0.7}
\begin{document}

\title{Scalable first-principles-informed quantum transport theory in two-dimensional materials}

\author{Sathwik Bharadwaj}
\affiliation{Department  of   Physics,  Worcester
	Polytechnic Institute, Worcester, Massachusetts 01609, USA.}
\affiliation{Center for Computational NanoScience,  Worcester
	Polytechnic Institute, Worcester, Massachusetts 01609, USA.}
\author{Ashwin Ramasubramaniam}
\email{ashwin@engin.umass.edu}
\affiliation{Department of Mechanical and Industrial Engineering, University of Massachusetts, Amherst, Massachusetts 01003, USA.}

\author{L. R. Ram-Mohan}
\email{lrram@wpi.edu}
\affiliation{Department  of   Physics,  Worcester
	Polytechnic Institute, Worcester, Massachusetts 01609, USA.}
\affiliation{Center for Computational NanoScience,  Worcester
	Polytechnic Institute, Worcester, Massachusetts 01609, USA.}
\vspace{0.2in}

\begin{abstract}
Accurate determination of carrier transport properties in two-dimensional (2D) materials is critical for designing high-performance nano-electronic devices and quantum information platforms. While first-principles calculations effectively determine the atomistic potentials associated with defects and impurities, they are ineffective for direct modeling of carrier transport properties at length scales relevant for device applications. Here, we develop a scalable first-principles-informed quantum transport theory to investigate the carrier transport properties of 2D  materials.  We derive a non-asymptotic quantum scattering framework to obtain transport properties in proximity to scattering centers. We then bridge our scattering framework with $\textit{\textbf{k}}\cdot\textit{\textbf{p}}$ perturbation theory, with inputs from first-principles electronic structure calculations, to construct a versatile multiscale formalism that enables modeling of realistic devices at the mesoscale. Our formalism also accounts for the crucial contributions of decaying evanescent modes across heterointerfaces. We apply this formalism to study electron transport in lateral transition-metal dichalcogenide (TMDC) heterostructures and show that material inclusions can lead to an enhancement in electron mobility by an order of magnitude larger than pristine TMDCs.
\end{abstract}
\maketitle

\begin{center}
\today
\end{center}
\section{Introduction}\label{sec:Intro}
Nanoelectronic devices, based on low-dimensional materials offers avenues to construct integrated circuits beyond the limits of Moore's law \cite{Li, Shalf, ashwin}. Atomically thin monolayers such as graphene \cite{Schwierz}, silicene \cite{Akinwande}, phospherene \cite{zhang_phospherene}, and transition metal dichalcogenides (TMDC) \mbox{\cite{Salahuddin, Cui, Bartolomeo}} have all been explored to construct high-performance field effect transistors, and these two-dimensional (2D) materials have shown great potential to succeed silicon in the next-generation computers. However, accurate prediction of transport properties in these materials is a crucial challenge that must be overcome before they can be used in practical devices \cite{Li}. 

Electron scattering due to imperfections, dopants/impurities, and phonons determine the carrier dynamics in 2D materials. Carrier transport properties can also be engineered by creating potential patterns and quantum confinements, such as through the application of local potentials via STM/AFM tips \cite{Levitov, Pasupathy}, surface functionalization by organic molecules \cite{ashwinNaveh, Margapoti, ulloa1}, or through spatial confinement in lateral heterostructures \cite{shenoy}. Density-functional theory (DFT) \cite{kohnsham} provides a parameter-free method for electronic structure calculations and the accurate determination of atomistic potentials associated with heterointerfaces, defects, and impurities. However, DFT is ineffective for direct
modeling of carrier transport properties at length scales relevant to device applications. There have been efforts to provide an atomistic quantum transport framework based on tight-binding methods \cite{kwant}, plane-wave representation of empirical pseudopotentials \cite{Vandenberghe}, and DFT-based non-equilibrium Green's function techniques \cite{Taylor, Brandbyge}. However, such simulations can become computationally expensive with increasing number of atoms. Moreover, scattering across the material interfaces have significant contributions from the decaying evanescent modes, which traditional scattering calculations do not account, since the probabilities of evanescent modes vanish in the asymptotic limit where the  boundary  conditions (BCs) are  applied  to  determine  the scattering amplitudes. Hence, we require a more detailed scattering framework where the scattering properties are evaluated in proximity to scattering centers.  

In this article, we present a framework for first-principles-informed continuum quantum transport theory. We develop a non-asymptotic quantum scattering theory, and show that bridging our scattering theory framework and the $\textit{\textbf{k}}\cdot\textit{\textbf{p}}$ perturbation theory, using inputs from \textit{ab-initio} electronic structure calculations provides a versatile multiscale formalism. In particular, we envision immediate applications of our theory in simulations and design of electron optics, and in nanoscale optoelectronic platforms. 

This article is organized as follows. In Sec.~\ref{sec:sourcesabsorbers}, we develop a non-asymptotic quantum scattering theory based on sources and absorbers. With this setup, we obtain the total wavefunction that includes both propagating and evanescent band contributions. This formalism is then extended to a multiband framework, by integrating with it the $\textit{\textbf{k}}\cdot\textit{\textbf{p}}$ perturbation theory using inputs from DFT calculations, as explained further in Sec.~\ref{sec:envelopetheory}. In that section we also explain the method to include thermal properties in our formalism through deformation potentials obtained via DFT calculations. In Sec.~\ref{sec:tmdc}, we apply our formalism to study transport properties in lateral TMDC heterostructures. We observe the emergence of Fano resonances in 2D materials with material inclusions. We also show the enhancement of electron mobility in lateral heterostructures comprised of combinations of TMDC monolayers. Concluding remarks are presented in Sec.~\ref{sec:conclusions}.   

\section{Construction of Sources and Absorbers}\label{sec:sourcesabsorbers}
\begin{figure}[ht]
\includegraphics[width=3in]{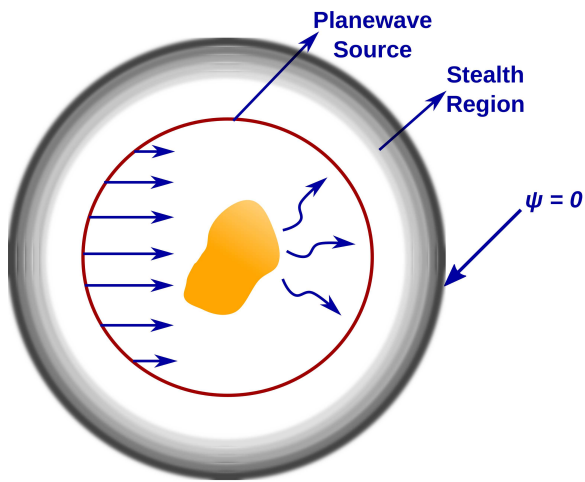}
\caption{Schematic representation of the construction of a circular source (red line) and absorber (gray region, also known as the stealth region) around a scattering center (green shaded region).  Dirichlet boundary conditions are applied at the boundary of the absorber. Amplitudes on the source circle are chosen as per Eq.~(\ref{eq:sourceamp}), so that the plane waves are injected into the region confining the scattering center.}
\label{fig:opendomain}
\end{figure}
The traditional treatment of scattering employs an {\it ad-hoc} condition that the distance between scattering centers and the observer
is asymptotically large. At large distances, the solution wavefunction is expressed as a linear combination of the incoming and outgoing basis functions with undetermined amplitudes. A partial wave analysis is then performed in the asymptotic limit to obtain these amplitudes in terms of the phase-shift parameters. For  example, in the 3D open  domain, the wavefunction is expressed in terms  of the spherical Hankel functions of the first kind  $h_m^{(1)}(kr)$ \cite{Adhikari1}. We can then easily deduce the scattering amplitudes since all Hankel functions have the  same  asymptotic form  given by  $e^{ikr}/r$, which represents the outgoing spherical wave \cite{sommerfeld} . Lord Rayleigh \cite{rayleigh} developed this method for the first time in  the context of  acoustic scattering, and later it was extended to quantum mechanics by Faxen and Holtsmark \cite{faxen}.   

In the asymptotic limit, only the radial component of the scattered current contributes to the cross-section. As a result, information about the angular current contributions \cite{cohenTannoudji}, and the distance between target and observer is lost in this analysis. In reports by Dai {\it et al.,}\cite{dai1,dai2} it has been shown that a rigorous scattering treatment without the asymptotic approximations introduces a modification factor to the scattering amplitude, and an additional phase factor to the scattered wavefunction. Hence, our first aim is to formulate a framework to obtain the scattering properties without applying asymptotic BCs. In the conventional variational formulation of the scattering theory, asymptotic BCs are applied either by mapping the far-field partial wave expansion into the near-field region \cite{porod} or by approximating the Sommerfeld radiation conditions \cite{sommerfeld,ludwig}. However, both of these approaches fail to account for the decaying evanescent solutions, since they inherit the asymptotic BCs.    
 
Previously, for electron scattering in waveguides, we proposed a method to reduce the scattering BCs to Dirichlet BCs by creating absorbers (stealth regions) on either end of the waveguide \cite{sathwik1}. An extension of this technique to 2D open domain is achieved here by creating circular absorbers around the scattering center, as shown in Fig.~\ref{fig:opendomain}. Within the absorber, we perform a coordinate transformation $\rho\rightarrow \rho\left(1+i\alpha(r)\right)$, and an energy transformation $E\rightarrow E\left(1+i\alpha(r)\right)$. Here, $\alpha(r)$ is the cubic Hermite interpolation polynomial varying smoothly from $0$ to $\alpha_{max}$.  Through such transformation, it has been shown that the no-reflection condition is satisfied \cite{sathwik1} and the absorber will not reflect any wave back into the active region confined within the absorber. Therefore, scattering properties in the active region remain unaffected by the presence of the absorber. As a result of such a transformation, the wavefunction decays rapidly within the absorber, and we can apply Dirichlet BCs at the boundary of the absorber. Through this technique, we can confine the computational domain, while preserving the near-field nature of the wavefunction in the region confined by the absorber (scattering region), that is otherwise lost due to the application of asymptotic BCs.

Once we enclose the active scattering region with the absorber, we
require a source (an electron antenna) in the active region to initiate the incoming plane waves (see Fig.~\ref{fig:opendomain}). This is achieved by introducing a source term in the Schrödinger equation. In order to derive this source term, we start with
the Green's function equation for the Schr\"odinger operator in the 2D circular coordinate system given by
\begin{align}\label{eq:schroedinger}
\left[\frac{1}{\rho}\frac{\partial}{\partial\rho}\left(\rho\frac{\partial}{\partial\rho}\right)+\frac{1}{\rho^2}\frac{\partial^2}{\partial\phi^2}+k^2\right]&\,G(\boldsymbol{\rho},\boldsymbol{\rho'})\nonumber\\ = S(\rho',\phi')\,&\frac{\delta\left(\rho-\rho'\right)}{\rho'}\,\delta\left(\phi-\phi'\right),
\end{align}
where $S(\rho',\phi')$ is the source term whose form is yet to be determined, the wavevector $k=\sqrt{{2m^* E}/{\hbar^2}}$, $E$ is the incoming energy, and $m^*$ is the effective mass. 
We expand the Green's function in the Fourier representations as
\begin{equation}\label{eq:greenfn}
G(\boldsymbol{\rho},\boldsymbol{\rho'}) = \frac{1}{2\pi}\sum_{m=-\infty}^{\infty} e^{im\left(\phi-\phi'\right)}g_m(\rho,\rho'). 
\end{equation} 
Inserting this Fourier expansion into Eq.~(\ref{eq:schroedinger}) we obtain 
\begin{align}
\frac{1}{2\pi}\sum_{m=-\infty}^{\infty}& e^{im\left(\phi-\phi'\right)}\left[\frac{1}{\rho}\frac{\partial}{\partial\rho}\left(\rho\frac{\partial g_m}{\partial\rho}\right)+\left(k^2-\frac{m^2}{\rho^2}\right)g_m\right]\label{eq:greensub1}\\ &= S(\rho',\phi')\,\frac{\delta\left(\rho-\rho'\right)}{\rho'}\,\delta\left(\phi-\phi'\right).\nonumber
\end{align}
Multiplying Eq.~(\ref{eq:greensub1}) by $e^{-im\phi}$ and integrating over $\phi$, from $0$ to $2\pi$, we obtain 
\begin{align}
\left[\frac{1}{\rho}\frac{\partial}{\partial\rho}\left(\rho\frac{\partial g_m}{\partial\rho}\right)+\left(k^2-\frac{m^2}{\rho^2}\right)g_m\right] = S(\rho',\phi')\,\frac{\delta\left(\rho-\rho'\right)}{\rho'}.
\end{align}
The radial part $g_m(\rho,\rho')$ is the Green's function for the one-dimensional Sturm Liouville operator \cite{MorseFeshbach} and will be of the form
\begin{equation}\label{eq:radialgreenfn}
g_m(\rho,\rho') = A_m\begin{cases}
J_m (k\rho) H_m(k\rho'),& \rho \leq \rho' \\
J_m (k\rho') H_m(k\rho),& \rho > \rho' 
\end{cases},
\end{equation}
where, $J_m(k\rho)$ and $H_m(k\rho)$ are the Bessel and Hankel function of first kind, respectively. 
We determine the coefficient $A_m$ by applying the jump condition at \mbox{$\rho = \rho'$} given by
\begin{align}
\frac{\partial g_m}{\partial \rho}\Biggr{|}_{\rho'+\epsilon}-\frac{\partial g_m}{\partial \rho}\Biggr{|}_{\rho'-\epsilon} &= \frac{S(\rho',\phi')}{\rho'},\nonumber\\
A_m\,\mathcal{W}\left[J_m(k\rho'),H_m(k\rho')\right] &= \frac{S(\rho',\phi')}{\rho'},
\end{align}
where, the Wronskian \mbox{$\mathcal{W}\left[J_m(k\rho'),H_m(k\rho')\right] = 2i/\pi\rho'$}. Thus,
\begin{equation}\label{eq:amcoeff}
A_m = -\frac{i\pi}{2}\,S(\rho',\phi').
\end{equation}
Substituting Eqs.~(\ref{eq:radialgreenfn}) and (\ref{eq:amcoeff}) into Eq.~(\ref{eq:greenfn}), we obtain the total Green's function of the form
\begin{align*}
G(\boldsymbol{\rho},\boldsymbol{\rho'}) &=  -\frac{i}{4}\,S(\rho',\phi')\times\nonumber\\&\sum_{m=-\infty}^{\infty} e^{im\left(\phi-\phi'\right)}\begin{cases}
J_m (k\rho) H_m(k\rho'),& \rho \leq \rho' \\
J_m (k\rho') H_m(k\rho),& \rho > \rho' 
\end{cases},\nonumber\\
\end{align*}
By using the addition theorem for Bessel functions \cite{MorseFeshbach} we have
\begin{align}
G(\boldsymbol{\rho},\boldsymbol{\rho'}) &= -\frac{i}{4}\,S(\rho',\phi')\,H_0(k\left|\boldsymbol{\rho}-\boldsymbol{\rho'}\right|).
\end{align}
We note that $H_0(k\left|\boldsymbol{\rho}-\boldsymbol{\rho'}\right|)$ represents the wave originating from the point source at $\boldsymbol{\rho'}$. As shown in Fig.~\ref{fig:opendomain}, to obtain a circular source we integrate $G(\boldsymbol{\rho},\boldsymbol{\rho'})$ over the angular coordinate $\phi'$ from $0$ to $2\pi$. Hence the wavefunction emerging from the circular source is given by
\begin{equation}\label{eq:circularwavefn}
\psi_{in}(\rho,\rho') = -\frac{i}{4}\int_{0}^{2\pi}d\phi' S(\rho',\phi')\,H_0(k\left|\boldsymbol{\rho}-\boldsymbol{\rho'}\right|).
\end{equation}
Our aim is to generate plane waves propagating in the forward direction from the circular source within the region $\rho\leq\rho'$. Hence, we choose the source term to be
\begin{equation}\label{eq:sourceamp}
S(\rho',\phi') = \frac{2i}{\pi}\sum_{n=-\infty}^{\infty} \frac{i^n\,e^{in\phi'}}{H_n(k\rho')}.
\end{equation}
Now, substituting Eq.~(\ref{eq:sourceamp}) into Eq.~(\ref{eq:circularwavefn}) we obtain
\begin{equation}
\psi_{in} =\sum_{m=-\infty}^{\infty} i^m\,e^{im\phi} \,\begin{cases}
\ J_m(k\rho)\hspace{0.55in},& \rho \leq \rho'; \\[6pt]
\displaystyle\frac{J_m(k\rho')}{H_m(k\rho')}H_m(k\rho),& \rho>\rho',
\end{cases}
\end{equation}
where we have utilized the orthogonal properties of the function $e^{in\phi'}$. We know the expansion of plane wave $e^{ikx} = e^{ik\rho\cos\phi} =\displaystyle\sum_{m=-\infty}^{\infty} e^{im\phi}i^m J_m(k\rho)$. Hence,
\begin{equation}
\hspace{-0.1in}\psi_{in} =\begin{cases}
\hspace{0.5in} e^{ikx}\hspace{1.14in},& \rho \leq \rho'; \\[6pt]
\displaystyle\sum_{m=-\infty}^{\infty} e^{im\phi} i^m\,\displaystyle\frac{J_m(k\rho')}{H_m(k\rho')}H_m(k\rho),& \rho>\rho',
\end{cases}
\end{equation}
and we obtain plane waves impinging on the scattering center from a circular source at $\rho =\rho'$, as shown in Fig.~\ref{fig:opendomain}. The wavefunction in the $\rho>\rho'$ region will simply be absorbed by the stealth region. 

Once we formulate the quantum scattering theory with the source and absorber, 
we can employ numerical methods to solve Eq.~(\ref{eq:schroedinger}) and obtain the total wavefunction $\psi$. In the presence of a scattering potential, we obtain the total wavefunction $\psi =\psi_{in} + \psi_{sc}$. Here, $\psi_{sc}$ includes both the radial and angular current contributions. In particular, $\psi_{sc}$ also includes the evanescent solutions in case of an absorbing scattering potential or in case of multiband scattering processes, as discussed in the next section.   
\begin{figure}
    \centering
    \includegraphics[width=3in]{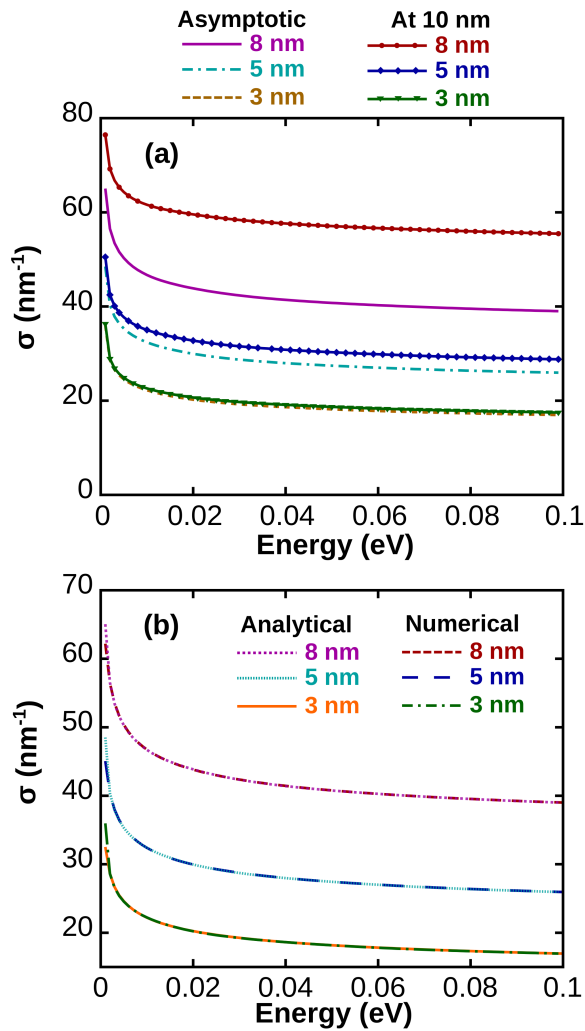}
    \caption{(a) The scattering cross-section length obtained through our scattering formalism is plotted as a function of the incoming energy. Here, the dotted line represents the cross-section measured by an observer at infinity, whereas the continuous line represents the cross-section measured by an observer at a finite distance of $10\,$nm. The radii of the hard circle potential considered here are $8\,$nm, $5\,$nm and  $3\,$nm.
    (b) The analytical result through partial wave analysis, and the numerical result obtained through our calculation for the scattering cross-section in a hard-circle potential is plotted as a function of energy.}
    \label{fig:hardcircle}
\end{figure}

Throughout this article, we have solved the Hamiltonian equation by casting it into an action integral. This action integral is solved using finite element analysis (FEA) \cite{ram_book,Zienkiewicz,jin,ramfem}. In FEA, we discretize the physical domain of interest into small elements. Within each element, we express the wavefunction as a linear combination of interpolation polynomials multiplied by undetermined coefficients. These coefficients correspond to the value of the wavefunction and their derivatives at the vertices (nodes) of the element. With this approach, one can systematically increase the accuracy through mesh size refinement ($h$-refinement) or by employing higher order interpolation polynomials ($p$-refinement).

To validate our scattering formalism, we consider the case of 2D scattering from a hard circle. The hard circle potential is given by
\begin{equation}
V(\rho) = \begin{cases}
0,& \rho \leq a\\
\infty,& \rho > a
\end{cases},
\end{equation}
where, $a$ is the radius of the circle. 
Through partial-wave analysis, we can obtain a closed-form result for the scattering cross-section length, $\sigma$, measured at the asymptotic distance (see Appendix~\ref{appdx:hardcircle}). In Fig.~\ref{fig:hardcircle}(b), we compare the analytical result (see Eq.~(\ref{eq:hardcirclesigmasum})), and the result obtained through our formalism. The behavior of $\sigma$ at low- and high-energy limits, observed in Fig.~\ref{fig:hardcircle}(b), is discussed in Appendix~\ref{appdx:hardcircle}. 

In Appendix~\ref{appdx:2Dscat}, we outline the scattering theory without applying asymptotic boundary conditions \cite{dai1,dai2}. In Fig.~\ref{fig:hardcircle}(a), we compare the cross-section length obtained by an observer at a very large distance (asymptotic limit) and by an observer at a finite distance of $10\,$nm,  and find a significant quantitative difference. In case of the hard circle potential, difference between the asymptotic and non-asymptotic predictions increases with decreasing distance between the observer and the hard circle.  
 In  Fig.~\ref{fig:hardcircle}(a), we see that the deviation between the prediction at the asymptotic limit, and for the observer at $10\,$nm is maximum for the hard circle radius of $8\,$nm. This is because the effective distance from the hard circle and the observer is least for the $8\,$nm potential, amongst the considered hard-circle radii. Thus confirming that in nanoscale systems, it is especially important to employ the non-asymptotic scattering theory developed here to obtain accurate transport properties. In Appendix~\ref{appdx:3Dscat}, we derive the source term for the non-asymptotic scattering theory in three-dimensions (3D). Such analysis in 3D can be employed to obtain transport properties in bulk materials.  

\section{Envelope function scattering theory with sources and absorbers}\label{sec:envelopetheory}
In multiband scattering processes, the wavefunction will have contributions from not only the propagating waves, but also from the exponentially decaying evanescent waves \cite{sathwik1}. This is particularly prevalent in transport across heterointerfaces. 
\begin{figure}[h]
	\includegraphics[width=3in]{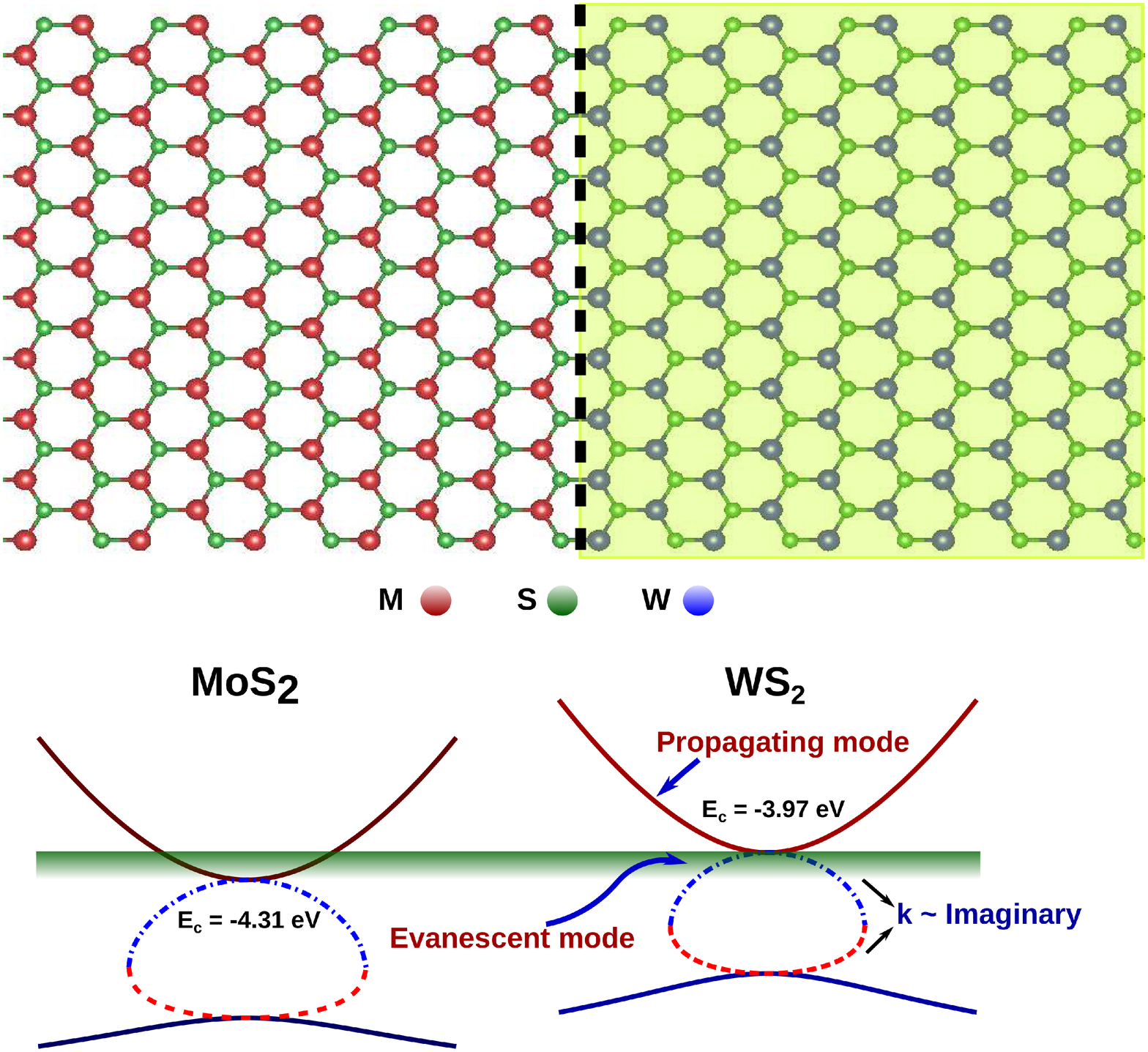}
	\caption{A schematic representation of the heterointerface between MoS$_2$ and WS$_2$ monolayers is shown.  We also display their corresponding conduction and valence bands near the $K-$point. MoS$_2$ \mbox{($E_c =-4.31\,$eV)} has a lower conduction band minimum than WS$_2$ \mbox{($E_c = -3.97\,$eV)}. Hence, below energies of $-3.97\,$eV, electron transport across the interface occurs only through evanescent modes.}
	\label{fig:mos2_ws2}
\end{figure}
For example, in Fig.~\ref{fig:mos2_ws2}, we consider a heterointerface formed between MoS$_2$ and WS$_2$ monolayers. From first-principles electronic structure calculations, we know that MoS$_2$ \mbox{($E_c =-4.31\,$eV)} has a lower conduction band minimum than WS$_2$ \mbox{($E_c = -3.97\,$eV)} monolayer. Hence, when an electron with energy $-4.31\,$eV$\leq\!\!E\!<\!\!-3.97\,$eV is injected from the MoS$_2$ to the WS$_2$ monolayer, transport can occur only through the evanescent modes. These evanescent modes are placed within the bandgap and will have purely imaginary wavevectors, which in turn results in exponentially decaying contributions to the scattered wavefunction. Traditional scattering calculations cannot account for these crucial contributions, since the probabilities of evanescent modes vanish at the asymptotic limit where the boundary conditions are applied to determine the scattering amplitudes. We therefore require a new quantum scattering framework as discussed in Sec.~\ref{sec:sourcesabsorbers}, that can accurately compute both the evanescent modes and angular current contributions in the near-field of the scattering center.  

The low-energy dynamics of charge carriers in semiconductors can be effectively described by the $\textit{\textbf{k}}\cdot\textit{\textbf{p}}$ Hamiltonian \cite{luttingerkohn, luttinger, bir_pikus, Dresselhaus}. This Hamiltonian provides an accurate characterization of the energy dispersion around the high-symmetry points of the Brillouin Zone, in terms of a small number of input band parameters. These band parameters, in turn, can be obtained from accurate first-principles calculations, for example, DFT or many-body perturbation theory \cite{C2DB}. In this section, we describe the multiband scattering theory using the envelope functions solutions of the $\textit{\textbf{k}}\cdot\textit{\textbf{p}}$ Hamiltonian. 

The $\textit{\textbf{k}}\cdot\textit{\textbf{p}}$ Hamiltonian effectively captures the coupled dynamics of the conduction and valence bands around the high-symmetry points of the Brillouin zone. This is also the region in which we are most interested in determining the transport properties. Effects of remote bands can be easily included by using L\"owdin perturbation theory \cite{lowdin}. Effects due to strain, magnetic field, and spin-orbit coupling can also be included as additional input parameters. Being a long-wavelength theory, $\textit{\textbf{k}}\cdot\textit{\textbf{p}}$ perturbation theory is well suited to simulate transport properties at device-relevant scales.  

Let us consider the Hamiltonian equation of the form
\begin{align}
\left(H_0 + V_0 +V_d-E\right)\cdot\psi_{nk} &\nonumber\\
= S(\rho',\phi')&\,\eta_{\rm scale}\,\mathbb{A}_{nk}\,\frac{\delta(\rho-\rho')}{\rho'},
\end{align}
where $S(\rho',\phi')$ is the source term defined in Eq.~(\ref{eq:sourceamp}), $\mathbb{A}_{nk}$ is the incoming amplitude, $n$ is the band-index, $k$ is the wavevector, and $\eta_{\rm scale}$ is a scale factor (derived in Appendix~\ref{appdx:scale}). Here, $V_0$ is the cell periodic potential, and $V_d$ is the defect/impurity potential obtained from DFT calculations. In the envelope-function approximation (EFA), the general form of the wavefunction and the incoming amplitude are considered as a linear combination of a finite number of bands \cite{luttingerkohn,kane1,kane2}, given by
\begin{align}
\psi_{nk} &= \sum^{{nband}}_{m=0} F_m\,u_{mk},\nonumber\\
\mathbb{A}_{nk} &= \sum^{{nband}}_{m=0} a_m\,u_{mk},
\end{align}
where $F_m$ is the slowly-varying envelope function, $u_{mk}$ is the cell-periodic Bloch function, $a_m$ is the incoming amplitude, and ${nband}$ is the number of bands considered. Here, the expansion is at a fixed $k$ value, and we will assume from now on that the expansion is around $k=0$. Within the framework of EFA, we perform `cell-averaging' by integrating over each unit cell of the crystal. Bloch functions satisfy Schrödinger’s equation with band-edge energies, and results in:
\begin{align}
\int_{\rm cell}d^2\rho\, u^{\dagger}_{n0}\cdot u_{m0} &= \delta_{mn},\nonumber\\
\int_{\rm cell}d^2\rho\, u^{\dagger}_{n0}\left(H_0+V_0\right)\cdot\,u_{m0} &= E_m\delta_{mn},\nonumber\\
\int_{\rm cell}d^2\rho\, u^{\dagger}_{n0}\,\textit{\textbf{p}}\cdot u_{m0} &= \textit{\textbf{p}}_{mn},
\end{align}
where $\textit{\textbf{p}}_{mn}$ is the momentum matrix element. This averaging procedure results in equation of the form
\begin{equation}\label{eq:kpsacteqn}
\hspace{-0.1in}\left[\left<\mathbf{H}_{kp}\right>_{\rm cell} +\left(V_d-E\right)\mathbf{1}\right]\cdot\mathbb{F} = S(\rho',\phi')\,\eta_{\rm scale}\,\mathbb{A}\,\frac{\delta(\rho-\rho')}{\rho'},
\end{equation}
where $\left<\mathbf{H}_{kp}\right>_{\rm cell}$ is the $\textit{\textbf{k}}\cdot\textit{\textbf{p}}$ Hamiltonian, and $\mathbf{1}$ is the identity matrix. From now on we denote $\left<\mathbf{H}_{kp}\right>_{\rm cell}$ simply as $\mathbf{H}_{kp}$. The envelope function is $\mathbb{F} = \left(\begin{array}{cc}
F_1     &  F_2\, ...
\end{array}\right)^{T}$, and the incoming amplitude is $\mathbb{A} = \left(\begin{array}{cc}
a_1     &  a_2\, ...
\end{array}\right)^{T}$.
In the absence of the defect potential $V_d$, the incoming amplitudes satisfy an equation of the form
\begin{equation}\label{eq:bulkkpeqn}
\left[\mathbf{H}_{kp} -E\,\mathbf{1}\right]\cdot\mathbb{A} = 0. 
\end{equation}
Hence, for a given energy, we can obtain $\mathbb{A}$ by solving the above eigenvalue equation. Note that Eq.~(\ref{eq:bulkkpeqn}) represents the standard $\textit{\textbf{k}}\cdot\textit{\textbf{p}}$ Hamiltonian equation for the pristine material. 

We cast Eq.~(\ref{eq:kpsacteqn}) into an action integral, as this allows us to solve the equation in the variational approach. The advantages of the action integral formulation are that there is no ambiguity about reordering of operators for symmetrization \cite{lrr_wavefunction} and, moreover we can easily evaluate the conserved current using the gauge-variational approach introduced by Gell-Mann and Levy \cite{gellmanlevy} and used extensively by us \cite{lrr_wavefunction} (see Appendix~(\ref{appdx:actionintegral})). This action integral is solved using finite element analysis. A detailed discussion of this variational numerical procedure is described in our earlier work \cite{sathwik1}.  

Within the Boltzmann transport formalism, important transport properties such as the mobility, conductance and Seebeck coefficients are expressed in terms of the total relaxation time $\tau(E)$. To determine $\tau(E)$, we need to consider both the intrinsic and extrinsic scattering rates. The extrinsic scattering rate arises from the potential scattering. According to Matthiessen’s law
\begin{equation}
\frac{1}{\tau(E)} = \frac{1}{\tau_e(E)} + \frac{1}{\tau_{ph}(E)},
\end{equation}
where, $\tau_e$ is the extrinsic carrier scattering time and $\tau_{ph}$ is the total intrinsic scattering time arising from the acoustic and optical phonon mode contributions. 

\subsection{Carrier scattering time}
Through kinectic theory \cite{liboff}, for elastic scattering processes in 2D, $\tau_e$ is given by
\begin{equation}
\tau_{e}(E) = \frac{1}{n_d\, \sigma_m \left<v\right>},
\end{equation}
where $n_d$ is the disorder density, $\left<v\right>$ is the average velocity, and $\sigma_m$ is the momentum scattering cross-section defined as
\begin{equation}
\sigma_m = \int_{0}^{2\pi} \frac{d\sigma}{d\phi}\left(1-\cos\phi\right)\,d\phi,
\end{equation}
where, ${d\sigma}/{d\phi}$ is the differential cross-section length. The average incident velocity is
\begin{equation}
\left<v\right> = \frac{1}{\pi}\int_{-\pi/2}^{\pi/2}\,d\beta\, v\,\cos{\beta},
\end{equation}
where \mbox{$v=\left|\nabla_k E_n(k)\right|/\hbar$} is the carrier velocity, and $\beta$ is the angle between the velocity vector and the longitudinal axis. For example, for a uniform velocity distribution, $\left<v\right> ={2v}/{\pi}$.  

\subsection{Phonon scattering time}
First-principles calculations provide a convenient way of determining the phonon scattering times by computing the deformation potentials of the crystal lattice \cite{Bardeen}. In 2D materials, the phonon scattering rate from the longitudinal and transverse acoustic phonons is given by \cite{Kaasbjerg, kim_eph, kim_eph_mos2}
\begin{equation}\label{eq:acousticph}
\frac{1}{\tau_{ph,0}} = \frac{m^{*}D_1^2\,k_B T}{\hbar^3\,\rho_{2D}\,c_s^2},
\end{equation}
where $m^{*}$ is the effective mass, $D_1$ is the first-order deformation potential, $T$ is the absolute temperature, $\rho_{2D}$ is the 2D ion mass density, and $c_s$ is the sound velocity of the acoustic phonons. The intervalley acoustic phonon and optical phonon rates are expressed in terms of the zero-order deformation potential $D_0$, and is given by
\begin{equation}\label{eq:opticalph}
\frac{1}{\tau_{ph,1}} = \frac{m^{*}D_0^2}{2\hbar^2 \omega_{\xi} \rho_{2D}}\left(e^{\hbar\omega_{\xi}/k_B T}\Theta(E-\hbar \omega_{\xi})+1\right)\,f^{BE}_{\xi},
\end{equation}
where $\Theta$ is the Heaviside step function, $\omega_\xi$ is the phonon frequency, and $f^{BE}_\xi$ is the Bose-Einstein distribution function for the phonon mode $\xi$. Effect of the Fr\"ohlich interaction can be implicitly added to the deformation potential \cite{kim_eph}. We neglect the first-order optical deformation potential contributions, that are typically small compared to the values from Eqs.~(\ref{eq:acousticph}) and (\ref{eq:opticalph}). Hence the total intrinsic phonon scattering rate is computed by summing over the all phonon modes, \mbox{${\tau_{ph}^{-1}} = \displaystyle\sum_{\xi} {\tau_{\xi}^{-1}}$}.

\begin{figure}
    \centering
    \includegraphics[width=3in]{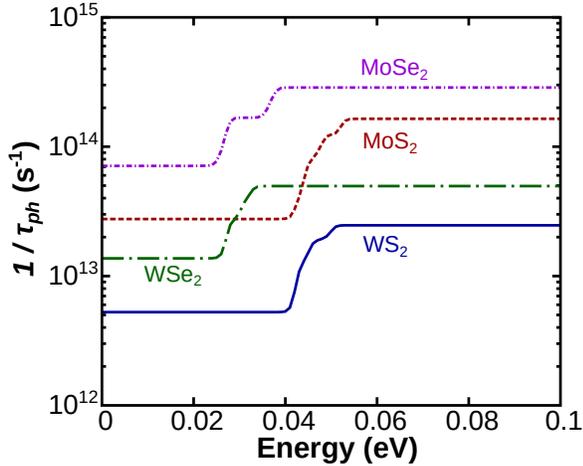}
    \caption{The total phonon scattering rates for the $K$-valley electrons are plotted as a function of energy for TMDCs at temperature $T= 300\,$K. Scattering rates are calculated using the deformation potentials listed in Ref.~\cite{Kaasbjerg, kim_eph, kim_eph_mos2}.}
    \label{fig:phononrate}
\end{figure}

In Fig.~\ref{fig:phononrate}, we have plotted the total intrinsic phonon scattering rate at room temperature for various TMDC monolayers. We have included deformation potentials for the acoustic and optical phonon modes corresponding to the transition, \mbox{$K\,\rightarrow\,\left\{K, K', Q, Q'\right\}$}. Emergence of contributions from optical phonon modes are observed as steps in the scattering rate.  We observe that MoSe$_2$ (WS$_2$) has the strongest (weakest) interaction with phonons for the considered TMDC monolayers. WX$_2$ has a greater electrical conductivity than MoX$_2$. These observations and scattering rate values are consistent with previous first-principles study in the literature \cite{Kaasbjerg, kim_eph}. 

\begin{figure*}[t]
	\includegraphics[scale=0.25]{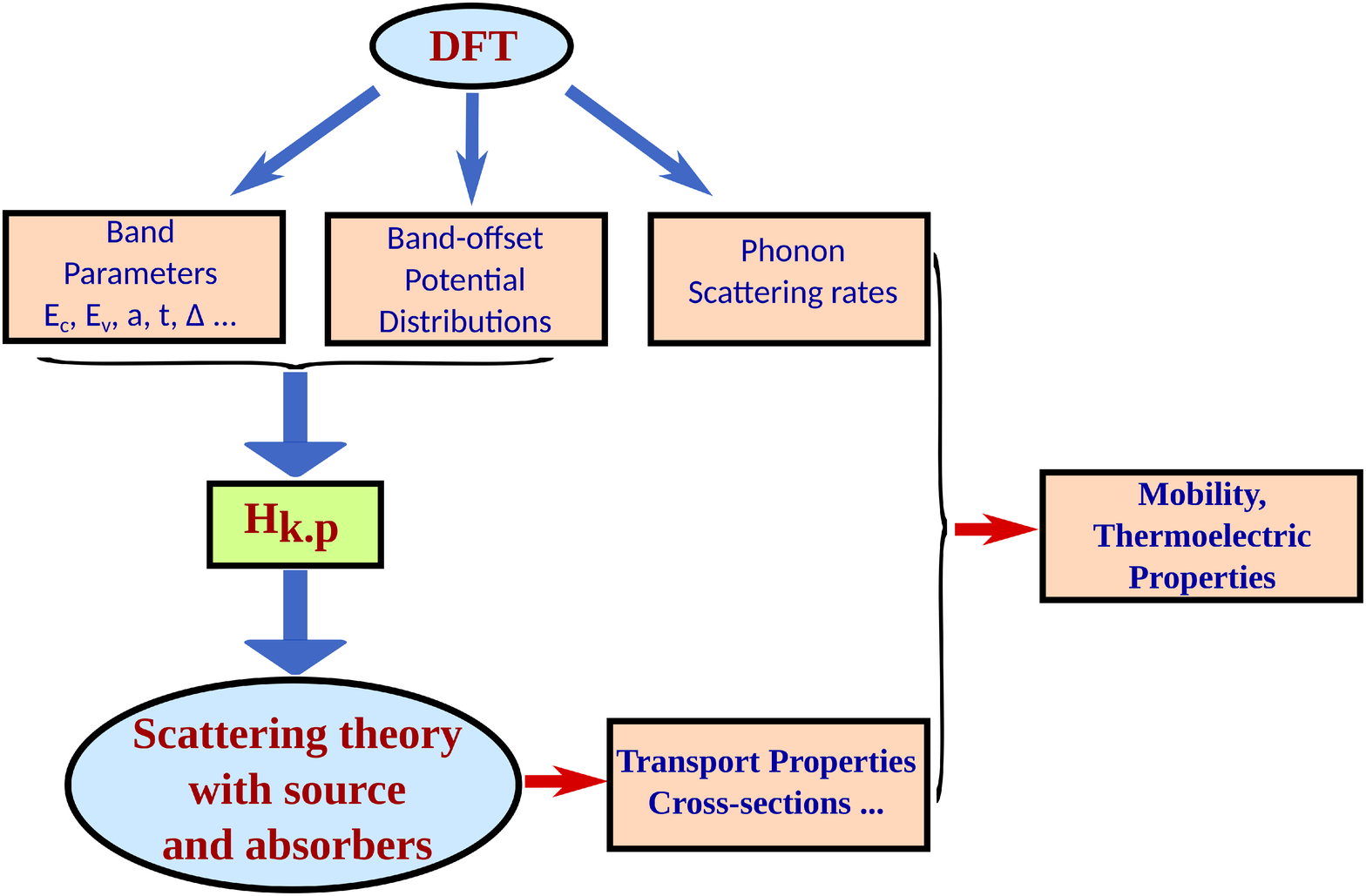}
	\caption{Flowchart of the quantum transport framework discussed in this article.}
	\label{fig:flowchart}
\end{figure*}

The flowchart displayed in Fig.~\ref{fig:flowchart} summarize the quantum transport framework discussed in this article. The band parameters, potential distributions, and phonon scattering rates are obtained through DFT calculations. These quantities will be an input to the $\textit{\textbf{k}}\cdot\textit{\textbf{p}}$ Hamiltonian which is then solved using the non-asymptotic quantum scattering theory. By combining the carrier scattering rate obtained through our calculations with the phonon scattering rate obtained though first-principles calculations, we can accurately obtain the carrier mobility and thermoelectric performance of the devices. In our theory, the scattering properties are calculated at a finite distance from the scattering center. Hence, it will accurately account for the evanescent mode contributions that are significant in scattering across heterointerfaces. This is evident from the enhancement of the electron mobility observed here, for TMDC monolayers with material inclusions. 
\section{Transport properties in TMDC heterostructures}\label{sec:tmdc}
Transition-metal dichalcogenides have attracted tremendous recent interest in fabricating 2D semiconductor nanoelectric circuits \cite{kis1,kis2,kis3}. In contrast to graphene, TMDCs host massive Dirac fermions due to their direct bandgap, which negates the probability of Klein tunneling \cite{ulloa_hardcirlceTMD}. Strong coupling between spin and valley degrees of freedom are observed in TMDC monolayers due to their lack of inversion symmetry \cite{cui1}. This makes them interesting candidates to realize valleytronic devices  as well \cite{ye}. 

Similar crystal structures and comparable lattice constants observed in MX$_2$ (M = Mo, W; X = S, Se) monolayers have motivated the study of TMDC-based quantum confinement in heterostructures \cite{shenoy, Kormanyos, Liu_pang, Wu_tong, Pawlowski, Pisoni, ulloa_ovando}. Experimentally, such lateral heterostructures can be realized through multistep chemical vapor deposition techniques, bottom up synthesis, and micromechanical techniques \cite{growth1, growth2, growth3, growth4, growth5, growth6, growth7, growth8, growth9, growth10}. The ability to pattern material inclusions in these systems provides an enhanced level of control over their electrical properties. Hence, they have great potential to realize in-plane transistors, photodiodes, cascade lasers, and CMOS circuits.

As shown in Fig.~\ref{fig:mos2_ws2}, the mismatch between band-offsets breaks the translational symmetry within planar TMDC heterostructures. Hence the evanescent modes (with purely imaginary wavevector) contribute heavily to their transport properties. Therefore, in 2D lateral heterostructures, the carrier scattering calculations have to carried out using the method described in Sec.~\ref{sec:envelopetheory}. In TMDCs, a seven-band $\textit{\textbf{k}}\cdot\textit{\textbf{p}}$ model can accurately capture the most important dispersion features of the conduction (CB) and valence (VB) bands \cite{kormanyos1}. We can apply L\"owdin perturbation theory \cite{lowdin} to reduce the seven band model into a two-band model corresponding to the lowest CB and highest VB. Here, we keep the terms up to second-order in the diagonal element in the Hamiltonian corresponding to the basis set \mbox{$\left\{\left|\phi_c,s, \eta\right>, \left|\phi_v,s, \eta\right>\right\}$}. Here, $s = \pm 1$ is the spin index, and $\eta = \pm 1$ denotes the valley $\pm K$. $C_{3h}$ symmetry dictates that the Hamiltonian should have the form 
\begin{equation}\label{eq:hkptmd}
H_{kp} = H_0 + at\left(\eta\, k_x \hat{\sigma}_x + i k_y \hat{\sigma}_y\right) - \lambda\,\eta \frac{(\hat{\sigma}_z - 1)}{2} s,
\end{equation}
where $\hat{\sigma}$ denotes the Pauli matrices, $a$ is the lattice constant, $t$ is the effective hopping integral, $\lambda$ is the spin-orbit (SO) parameter, and $H_0$ is given by
\begin{equation}
H_0 = \left[\begin{array}{cc}
   E_c + \alpha_{s} k^2  & 0 \\
   0   &  E_v +\beta_{s} k^2
\end{array}\right].
\end{equation}
Here, $\alpha_{s}, \beta_{s}$ are the material parameters, and $E_c$ and $E_v$ are the CB minimum and the VB maximum, respectively. As the $\textit{\textbf{k}}\cdot\textit{\textbf{p}}$ theory is corrupted with the occurrence of spurious solutions, we have utilized the Foreman transformations \cite{foreman} to eliminate the spurious solutions, and hence, we set $\beta_s = 0$. In Table~\ref{tab:parameters} we have listed all the \mbox{$\textit{\textbf{k}}\cdot\textit{\textbf{p}}$} parameters used in our calculations for MoS$_2$, WS$_2$, MoSe$_2$, WSe$_2$ monolayers. For now we have neglected strain matrix elements since MoS$_2$ (MoSe$_2$) and WS$_2$ (WSe$_2$) have commensurate lattice constants. Material parameters between the two layers are smoothly interpolated through the self-consistent potential distribution obtained from DFT calculations.
\begin{table}
	   \caption{The $\textit{\textbf{k}}\cdot\textit{\textbf{p}}$ parameters used in our calculations are listed. These parameters were obtained from the previously reported first-principles study \cite{C2DB, kormanyos1}. }
	\label{tab:my_label}
    \centering
   \begin{tabular}{c c c c c c c}
      \hline\hline & & & & & & \\
      & $E_c$  & $E_v$ & $a$  & $t$ & $\lambda$ & $\alpha_{+}(\alpha_{-})$\\[5pt]
      & (eV) & (eV) & (\angstrom) & (eV) & (eV) & (eV$\cdot$\angstrom$^2$) \\[5pt]
      \hline
      & & & & & & \\
      MoS$_2$ &-4.31 &-5.89 &3.184 &1.059 &0.073 &-5.97(-6.43) \\[5pt]
      WS$_2$ &-5.97 &-5.50 &3.186 &1.075 &0.211 &-6.14(-7.95) \\[5pt]
      MoSe$_2$ &-3.91 &-5.23 &3.283 &0.940 &0.090 &-5.34(-5.71) \\[5pt]
      WSe$_2$ &-3.61 &-4.85 &3.297 &1.190 &0.230 &-5.25(-6.93) \\[5pt]
      \hline\hline
   \end{tabular}\label{tab:parameters}
\end{table}
\begin{figure}
    \centering
    \includegraphics[width=3in]{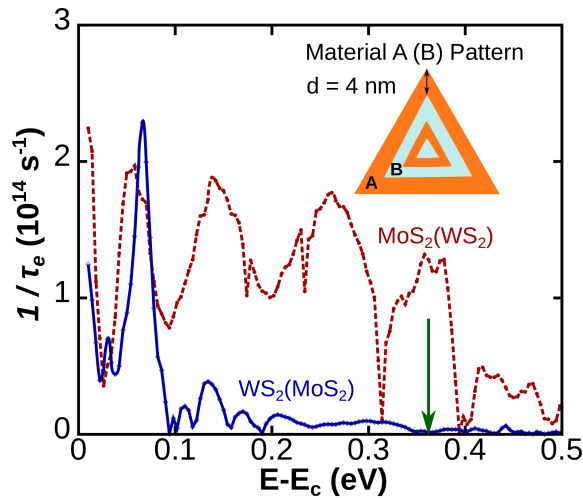}
    \caption{Electron scattering rate is plotted as a function of energy for MoS$_2$(WS$_2$) and WS$_2$(MoS$_2$) triangular heterostructures.  A(B) material pattern is displayed in the inset. The material parameters employed here are listed in Table~\ref{tab:parameters}, and $n_d = 10^{12}\,$cm$^{-2}$. The width of each shaded strip is taken to be 4\,nm. Scattering rates are calculated using the envelope-function scattering theory discussed in Sec.~\ref{sec:envelopetheory}.}
    \label{fig:lenseTMD}
\end{figure}

\subsection{Fano Resonance}
In Fig.~\ref{fig:lenseTMD}, we have plotted the electron scattering rate for a 2D triangular superlattice between MoS$_2$ and WS$_2$ monolayers. There have been experimental efforts to realize such structures through lateral epitaxy \cite{growth9}. MoS$_2$ has a lower CB minimum as compared to WS$_2$. Hence, below $E < -3.97\,$eV (marked by a green arrow in Fig.~\ref{fig:lenseTMD}) transport in WS$_2$ layers can occur only through the evanescent modes. Evanescent wavevectors are the purely imaginary $k$ roots of the $\textit{\textbf{k}}\cdot\textit{\textbf{p}}$ Hamiltonian defined in Eq.~(\ref{eq:hkptmd}). Firstly, we observe that the WS$_2$ (MoS$_2$) structure has a lower scattering rate (higher lifetime) as compared to the MoS$_2$ (WS$_2$) structure. This is because an electron injected from the WS$_2$ layer can scatter through the additional propagating conduction channel available from MoS$_2$ inclusions. For transport in the MoS$_2$ (WS$_2$) structure, we observe that at the incoming energies $E>-3.97\,$eV ($E-E_c = 0.34\,$eV, marked by an arrow in Fig.~\ref{fig:lenseTMD}), corresponding to the CB minimum of WS$_2$, the electron scattering rate decreases significantly, since the evanescent modes disappear above this energy. The resonances observed in the MoS$_2$ (WS$_2$) structure, below the energy $E<-3.97\,$eV are due to the interaction with the trapped electron states within the WS$_2$ barrier. 

\begin{figure}
    \centering
    \includegraphics[width = 3in]{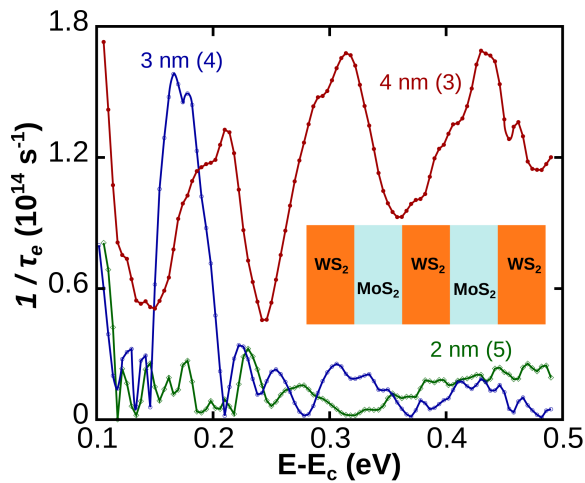}
    \caption{Electron scattering rate is plotted as a function of energy for  WS$_2$-MoS$_2$ lateral heterostructures.  Electrons are initiated from the WS$_2$ layer. Here, the notation 4\,nm(3) represents that there are 3 periods of WS$_2$ and MoS$_2$ layers, and each layer has a width 4\,nm. The material parameters employed here are listed in Table~\ref{tab:parameters} and $n_d = 10^{12}\,$cm$^{-2}$. Scattering rates are calculated using the envelope function scattering theory discussed in Sec.~\ref{sec:envelopetheory}.}
    \label{fig:strip}
\end{figure}

For transport in the WS$_2$ (MoS$_2$) structure, we observe both a maximum and minimum in the electron scattering rate around $E-E_c = 0.1\,$eV (see Fig.~\ref{fig:lenseTMD}). This marks the occurrence of a Fano resonance in the transmission profile, an effect analogous to that of atomic autoionization \cite{Fano}, first observed in the context of inelastic electron scattering by a helium atom. Analogous Fano profiles are observed around the subband minima of a quantum waveguide with attractive potentials \cite{sathwik2}. Here, the MoS$_2$ inclusion acts as a quantum well and leads to the formation of quasi-bound metastable states. Interaction between the propagating modes and the quasi-bound states leads to the Fano resonance. Multiple Fano resonances are observed when we have a superlattice of WS$_2$ and MoS$_2$ layers as shown in Fig.~\ref{fig:strip}. Formation of Fano resonances leads to an enhancement in the thermoelectric figure-of-merit far beyond the pure 2D monolayers. This phenomenon will be discussed in detail elsewhere \cite{sathwik_thermoelectric}.    

\subsection{Enhancement of Electron Mobility} 
In this section, we study the electron mobility in TMDC heterostructures. Around the $K$-valley, the CB is isotropic and parabolic in nature. Hence, the mobility can be defined as
 \begin{equation}
\mu = \frac{e}{m^{*}}\frac{1}{n}\int dE\,g(E)\,E\,\tau(E)\left(-\frac{\partial f(E)}{\partial E}\right),
 \end{equation}
where, the 2D carrier density $n=\int dE\,g(E)\,f(E)$, $g(E)$ is the density of states, and the Fermi-Dirac distribution is \mbox{$f(E)=\left(1+\exp\left[(E-E_F)/k_B T\right]\right)^{-1}$}. Due to the parabolic nature of CB, we consider constant \mbox{$g(E) = g_s g_v m_K^{*}/2\pi\hbar^2$}, where $g_s$ and $g_v$ \mbox{($g_v = 1, 2$ for VB, CB)} are the spin and valley  degeneracies, respectively.

\begin{figure}
    \centering
    \includegraphics[width=3in]{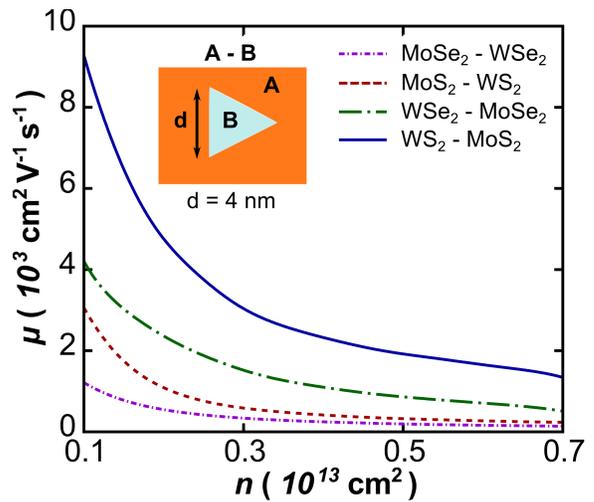}
    \caption{Electron mobility at room temperature as a function of 2D carrier density for the triangular inclusions in TMDC heterostructures. The material parameters employed here are listed in Table~\ref{tab:parameters} and $n_d = 10^{12}\,$cm$^{-2}$.}
    \label{fig:mobility}
\end{figure}

In Fig.~\ref{fig:mobility}, we plot the total electron mobility (including both phonon and electron contributions) at room temperature for the $n$-type TMDC monolayers with triangular inclusions. The density of triangular inclusions is considered to be $n_d = 10^{12}\,$cm$^{-2}$. We observe an enhancement in the mobility at low carrier density. The WS$_2$ monolayer with MoS$_2$ inclusion has the highest mobility amongst the considered combinations. This can be attributed to two main factors: 1) additional CB channels available from MoS$_2$ inclusions enhance the electron scattering lifetime, 2) WS$_2$ monolayer has the highest phonon scattering lifetime amongst the family of TMDC monolayers (see Fig.~\ref{fig:phononrate}). The MoS$_2$ monolayer with WS$_2$ inclusion, on the other hand has lower mobility. This is because the evanescent modes, being real functions, will not contribute to the probability current, thereby suppressing the overall mobility. We also note that the mobility observed here in WX$_2$ monolayers with triangular inclusions is an order of magnitude larger than the phonon-limited mobility and the mobility calculated in layers with charged vacancies. In charged vacancies, Coulomb contributions suppress the mobility \cite{tonylow}. Hence, short-range potentials such as the inclusions considered here are very promising candidates to realize high-mobility devices.  

\section{Conclusions}\label{sec:conclusions}
In this article, we have developed a non-asymptotic quantum scattering theory using a novel method of sources and absorbers. The Cauchy boundary conditions that are necessary in the variational formulation of scattering are reduced to simpler Dirichlet boundary conditions by constructing a perfect absorber (stealth region) around the scattering centers. This allows us to define a finite computational domain. In the conventional scattering theory, scattering properties are determined at the asymptotic distance, resulting in the loss of evanescent mode contributions in multiband scattering processes and in the presence of absorbing scattering potentials. Introducing an absorber facilitates the evaluation of the scattering properties at a finite distance, and preserves the distance information between the observer and the scattering center. Once we enclose the active scattering region with the absorber, we require a carrier source to generate plane-waves and initiate the scattering event. We have derived the expression for the circular source term in Eq.~(\ref{eq:sourceamp}) introduced in the right-hand side of the Hamiltonian equation. 

We have shown that even for the simplest case of scattering from a hard circle potential, the cross-section length measured at a finite distance using our formalism, and at the asymptotic distance obtained through the analytical formulation show significant quantitative differences. As the carrier scattering time is inversely proportional to the cross-section length, our formalism is of immediate relevance in nanoscale systems, where transport measurements are made in proximity to the scattering centers.     

We have shown that bridging our scattering theory framework with the $\textit{\textbf{k}}\cdot\textit{\textbf{p}}$ perturbation theory, with inputs from \textit{ab-initio} electronic structure calculations, provides a versatile multiscale formalism. The continuum nature of our formalism enables the modeling of realistic devices, scaling from hundreds to thousands of atoms. Carrier scattering rates obtained through our formalism combined with the phonon scattering rates obtained through DFT calculations can accurately calculate the thermoelectric properties of the nano-devices. 

We obtained the phonon scattering rates for MX$_2$ \mbox{(M = Mo, W; X = S, Se)} monolayers through deformation potential calculations. We observed that the WS$_2$ (MoSe$_2$) has the highest (lowest) phonon scattering time in the family of MX$_2$ monolayers. This observation is consistent with other first-principles study in the literature. 
As an application of our formalism, we studied the transport properties in lateral TMDC heterostructures. Material inclusions in TMDCs acts as a short-range scattering centers. We observed the emergence of novel Fano resonances for the first time in 2D materials, when MoS$_2$ is encapsulated within WS$_2$ monolayer. The MoS$_2$ inclusion here acts as a quantum well, and forms the quasi-bound states that interact with the propagating modes leading to these resonances. 

Lastly, we studied mobility of electrons as a function of carrier density in a family of TMDC layers with triangular inclusions. Here, WS$_2$ monolayer with MoS$_2$ inclusions is observed to have the highest electron-mobility, which is by an order of magnitude larger than the phonon-limited mobility. Hence, such lateral TMDC heterostructures should be explored as candidates to realize high-mobility devices.   


\begin{acknowledgments}
Calculations presented here were performed using computational resources supported by the Academic and Research Computing Group at WPI.
\end{acknowledgments}
\appendix
\section{Scattering cross-section length at a finite distance}\label{appdx:2Dscat}
The differential cross-section is determined by dividing the scattered flux by the incoming flux, given by
\begin{equation}
\frac{d\sigma}{d\phi} = \frac{{\textbf{J}^{\rm sc}\cdot d\mathbf{S}}}{{{J}_{\rm in}}},
\end{equation}
where $\textbf{J}^{\rm sc}$ is the scattered current, and $J_{\rm in} = \hbar k/ m^*$ is the magnitude of the current from the incoming plane wave. In the traditional scattering theory, the detector is at infinity, and the outgoing wave in 2D is a circular wave of the form $e^{ik\rho}/\sqrt{\rho}$. When the observer is at a finite distance, $\rho$, the outgoing wave-front incident on the observer will have both radial and angular components. The surface element $d\mathbf{S}$ is considered to be orthogonal to the outgoing current \cite{dai1,dai2}. In a 2D scattering,
\begin{align}
{\textbf{J}^{\rm sc}\cdot d\mathbf{S}} &= J^{\rm sc}\mathbf{n}\cdot dS\mathbf{n},\nonumber\\
&= J^{\rm sc}\sqrt{g}d\phi,
\end{align}
where $g$ is the determinant of the metric $g_{\mu\nu}$. In 2D circular coordinate system
\begin{equation}
\sqrt{g} = \rho\,\sqrt{1+\frac{1}{\rho^2}\left(\frac{d\rho}{d\phi}\right)^2}.
\end{equation}
Hence,
\begin{align}\label{eq:scatcurrent}
{\textbf{J}^{\rm sc}\cdot d\mathbf{S}} &= J^{\rm sc}\sqrt{1+\frac{1}{\rho^2}\,\left(\frac{d\rho}{d\phi}\right)^2}\rho\,d\phi,
\end{align}
where
\begin{align}
J^{\rm sc}_{\rho} &= \displaystyle\frac{\hbar}{m}\,{\rm Im}\left(\psi^{\dagger}_{sc}\frac{\partial \psi_{sc}}{\partial\rho}\right),\nonumber\\
J^{\rm sc}_{\phi} &= \displaystyle\frac{\hbar}{m}\,{\rm Im}\left(\frac{\psi^{\dagger}_{sc}}{\rho}\frac{\partial \psi_{sc}}{\partial\phi}\right).\nonumber
\end{align}

In order to compare with the expression given in the standard scattering theory, we would like to express the scattered flux in Eq.~(\ref{eq:scatcurrent}) only in terms of the scattering current components. We know that
\begin{align}
\frac{\partial}{\partial x} &= \cos\phi \frac{\partial}{\partial r} - \frac{\sin\phi}{r}\frac{\partial}{\partial \phi},\nonumber\\
\frac{\partial}{\partial y} &= \sin\phi \frac{\partial}{\partial r} + \frac{\cos\phi}{r}\frac{\partial}{\partial \phi}.
\end{align}

Hence we can express the scattered current as
\begin{align}\label{eq:scatcomp}
\mathbf{J^{\textrm{sc}}} =\left[\begin{array}{c}
J^{\textrm{sc}}_x \\ J^{\textrm{sc}}_{y} 
\end{array}	\right] = \left[\begin{array}{cc}
\cos\phi & -\sin\phi  \\
\sin\phi &  \cos\phi  
\end{array}\right]\cdot \left[\begin{array}{c}
J^{\textrm{sc}}_\rho \\ J^{\textrm{sc}}_{\phi} 
\end{array}	\right].
\end{align}
Now let us take a surface of revolution for the scattered current, expressed in a parametric form  $\rho = \rho\left(\phi\right)$. The tangential vector at a point on this surface is 
\begin{equation}
\mathbf{t} = \left[\begin{array}{c}
-\rho(\phi)\sin\phi \\ \,\rho(\phi)\cos\phi
\end{array}\right],
\end{equation}
and the normal vector is
\begin{align}
\mathbf{n} = \frac{\partial \mathbf{t}}{\partial \phi} = \left[\def\arraystretch{2.0}\begin{array}{c}
-\rho(\phi)\cos\phi - \displaystyle\frac{\partial \rho(\phi)}{\partial \phi}\sin\phi \\
-\rho(\phi)\sin\phi + \displaystyle\frac{\partial \rho(\phi)}{\partial \phi}\cos\phi 
\end{array}\right].
\end{align}
Hence, the scattered current is given by
\begin{align}\label{eq:scatcomptot}
\mathbf{J^{\textrm{sc}}} &= 
J^{\textrm{sc}}\,\frac{\mathbf{n}}{\left|\mathbf{n}\right|},\\ &= \frac{J^{\textrm{sc}}}{\sqrt{\rho^2+ \left(\displaystyle\frac{\partial\rho}{\partial\phi}\right)^2}}\,\def\arraystretch{2.0}\left[\begin{array}{c}
-\rho(\phi)\cos\phi - \displaystyle\frac{\partial \rho(\phi)}{\partial \phi}\sin\phi \\
-\rho(\phi)\sin\phi + \displaystyle\frac{\partial \rho(\phi)}{\partial \phi}\cos\phi 
\end{array}\right].\nonumber
\end{align}
Equating Eq.~(\ref{eq:scatcomp}) and (\ref{eq:scatcomptot}), we obtain the angular components of the scattered as
\begin{align}
J^{\textrm{sc}}_\rho &= \frac{-\rho J^{\textrm{sc}}}{\sqrt{\rho^2+ \left(\displaystyle\frac{\partial\rho}{\partial\phi}\right)^2}}, \nonumber\\[8pt]
J^{\textrm{sc}}_\phi &= \frac{\displaystyle\frac{\partial \rho}{\partial \phi} J^{\textrm{sc}}}{\sqrt{\rho^2+ \left(\displaystyle\frac{\partial\rho}{\partial\phi}\right)^2}}.
\end{align} 
Hence we find that 
\begin{equation}
\frac{J^{\textrm{sc}}_{\phi}}{J^{\textrm{sc}}_{\rho}} = -\frac{1}{\rho}\,\frac{d\rho}{d\phi}. 
\end{equation}
By substituting this relation in Eq.~(\ref{eq:scatcurrent}), we obtain
\begin{align}\label{eq:fluxexp1}
{\textbf{J}^{\textrm{sc}}\cdot d\mathbf{S}} &= J^{\textrm{sc}}\sqrt{1+\left(\frac{J^{\textrm{sc}}_{\phi}}{J^{\textrm{sc}}_{\rho}}\right)^2}\rho\,d\phi.
\end{align}
We know that $J^{\textrm{sc}} = \sqrt{\left(J^{\textrm{sc}}_\rho\right)^2 + \left(J^{\textrm{sc}}_\theta\right)^2} = J^{\textrm{sc}}_{\rho}\,\left[{1+\left({J^{\textrm{sc}}_{\phi}}/{J^{\textrm{sc}}_{\rho}}\right)^2}\right]$. 
Substituting this relation to Eq.~(\ref{eq:scatcurrent}) we obtain
\begin{align}
{\textbf{J}^{\rm sc}\cdot d\mathbf{S}} 
&= J^{\rm sc}_{\rho}\,\left[{1+\left(\frac{J^{\rm sc}_{\phi}}{J^{\rm sc}_{\rho}}\right)^2}\right]\rho\,d\phi,\nonumber
\end{align}
and the differential cross-section length is given by
\begin{equation}
\frac{d\sigma}{d\phi} = \frac{J^{\rm sc}_{\rho}\,\left[{1+\left(\displaystyle\frac{J^{\rm sc}_{\phi}}{J^{\rm sc}_{\rho}}\right)^2}\right]\rho}{J_{\rm in}}.
\end{equation}
Note that as the source and observer are pushed to infinity $(\rho \rightarrow \infty)$, 
\begin{align}
J^{\rm sc}_{\phi}\,\rightarrow\,& 0,\nonumber\\
J^{\rm sc}\,\rightarrow\,& \frac{\hbar}{m^*}\frac{\left|f\right|^2}{\rho},\\
\frac{d\sigma}{d\phi }\,\rightarrow\,&  \frac{\left|f\right|^2}{k},\nonumber
\end{align}
where $f$ is scattering amplitude. Hence, we obtain the standard definition of the differential cross-section length \cite{Adhikari1} in the limit $\rho\rightarrow\infty$.

\section{Scattering from a Hard circle potential}\label{appdx:hardcircle}
In this appendix, we discuss the analytical results for the scattering from a hard circle potential in 2D. Even though, this problem has been addressed before in the literature \cite{Lapidus, Shertzer}, the limiting behavior has not been discussed.  

The hard circle potential of radius $a$ in a 2D space is given by
\begin{equation}
V = \begin{cases}
\infty,& \rho \leq a \\
0,& \rho > a
\end{cases}.
\end{equation}
The solution wavefunction for the above potential is of the form
\begin{equation}
\psi (\rho, \phi) = \begin{cases}
0, & \rho \leq a \\
e^{ik\rho\,\cos\phi}+\displaystyle\sum_{n=-\infty}^{\infty}c_n\,e^{in\phi}H_n(k\rho),& \rho > a
\end{cases}.
\end{equation}
To determine the coefficients $c_n$, we utilize the boundary condition $\psi(a, \phi) = 0$. Hence,
\begin{equation}
c_n = \frac{-i^n J_n(ka)}{H_n(ka)},
\end{equation}
where we have used the expansion of the planewave \mbox{$e^{ik\rho cos\phi} =\displaystyle\sum_{n=-\infty}^{\infty} (i^n)e^{in\phi} J_n(k\rho)$}. Through partial-wave analysis, we obtain the total cross-section length given by
\begin{equation}\label{eq:hardcirclesigmasum}
\sigma = \frac{4}{k}\,\displaystyle\sum_{n=-\infty}^{\infty}\frac{\left|J_n(ka)\right|^2}{\left|H_n(ka)\right|^2}.
\end{equation}
To compare the quantum mechanical calculations with classical mechanics predictions, we consider the cross-section length in the high energy limit $ka >> 1$. We know that,
\begin{align}
J_n(ka) &\xrightarrow[ka \rightarrow \infty]{} \sqrt{\frac{2}{\pi k a}}\cos\left(ka-\frac{n\pi}{2}-\frac{\pi}{4}\right),\nonumber\\
H_n(ka) &\xrightarrow[ka \rightarrow \infty]{} \sqrt{\frac{2}{\pi k a}}\ \,e^{\displaystyle i\left(ka-\frac{n\pi}{2}-\frac{\pi}{4}\right)}.
\end{align}
Therefore,
\begin{equation}\label{eq:hardcirclesigma}
\sigma \simeq \frac{4}{k}\,\displaystyle\sum_{n=-\infty}^{\infty}\cos^2\left(ka-\frac{n\pi}{2}-\frac{\pi}{4}\right).
\end{equation}
Classically, the particle will not be deflected by the potential if the impact parameter $b > a$. The angular momentum $L_\phi$ will follow the limit $\left|L_\phi\right| = n\hbar \leq \hbar ka$. Hence, we can restrict the sum in Eq.~(\ref{eq:hardcirclesigma}) between the limits $-ka \leq n \leq ka$. With this simplification, we obtain the cross-section length
\begin{equation}
\sigma \simeq 4a, 
\end{equation}
which is twice the classical mechanics prediction \cite{Lapidus}. 

To obtain the low energy limit ($ka << 1$) of the cross-section length, we retain only the first term ($n = 0$) in the series given in Eq.~(\ref{eq:hardcirclesigmasum}). In this limit,
\begin{align}
J_0(ka) &\xrightarrow[ka \rightarrow 0]{} 1,\nonumber\\
H_0(ka) &\xrightarrow[ka \rightarrow 0]{} \frac{2}{\pi} \ln{(ka)}.
\end{align}
Hence,
\begin{equation}
\sigma \simeq \frac{\pi^2}{k} \frac{1}{\left(\ln{(ka)}\right)^2} .
\end{equation}
This is not an analytic function. Therefore, as $k\rightarrow 0$, cross-section length will diverge even though the scattering current vanishes.

\section{Scattering theory in three-dimensions using sources and absorbers}\label{appdx:3Dscat}
In this appendix, we set up the framework for the quantum scattering theory in 3D, using our source and absorber scheme. In 3D, we consider a spherical source, and the absorber will be a spherical shell. An incoming plane wave will be of the form
\begin{equation}
e^{i\mathbf{k}\cdot\mathbf{r}} = 4\pi\,\sum_{l=0}^{\infty}\,\sum_{m=-l}^{l} i^l\,Y_{ml}^{*}(k_\theta,k_\phi)Y_{ml}(\theta,\phi)j_l(kr),
\end{equation}
where, $Y_{ml}$ are the Laplace spherical harmonics, $j_l(kr)$ are the spherical Bessel function, and $(k_\theta,k_\phi)$ define the direction of the wavevector $\mathbf{k}$. Following a similar scheme as in Sec.~\ref{sec:sourcesabsorbers}, we can derive the corresponding source term 
\begin{equation}
S(r',\theta',\phi') = \frac{4\pi i}{k^2}\,\sum_{l=0}^{\infty}\,\sum_{m=-l}^{l} \frac{i^l\,Y_{ml}^{*}(k_\theta,k_\phi)Y_{ml}(\theta',\phi')}{h_l(kr')},
\end{equation}
where, $h_l(kr')$ is the spherical Hankel function. With this set up the incoming wavefunction in the absence of external potential is given by
\begin{equation}\nonumber
\hspace{-0.1in}\psi_{in} =\begin{cases}
\hspace{0.5in} e^{i\mathbf{k}\cdot\mathbf{r}}\hspace{1.14in}&, r \leq r' ; \\[6pt]
 {4\pi }\displaystyle\sum_{l=0}^{\infty}\sum_{m=-l}^{l} {i^l\,Y_{ml}^{*}(k_\theta,k_\phi)Y_{ml}(\theta',\phi')}\!&\!\hspace{-0.3cm}\displaystyle\frac{j_{l}(kr')h_l(kr)}{h_l(kr')}\\&, r > r' .
\end{cases}
\end{equation}
Hence, we obtain plane waves incident on the scattering center from a spherical source at $r =r'$. 

\section{Action integral formation}\label{appdx:actionintegral}
In this appendix, we explain the action integral formulation employed to solve Eq.~(\ref{eq:kpsacteqn}). Let us consider a general form of the $\textit{\textbf{k}}\cdot\textit{\textbf{p}}$ Hamiltonian in a 2D material given by
\begin{equation}
\mathbf{H}_{kp} = \mathbf{A}_{xx} k_x^2 + \mathbf{A}_{yy}k_y^2 +\mathbf{A}_{xy}k_{x}k_{y} + \mathbf{B}_{x}k_x + \mathbf{B}_{y}k_y + \mathbf{C},
\end{equation}
where $\mathbf{A}_{xx}, \mathbf{A}_{yy}, \mathbf{A}_{xy}, \mathbf{B}_x, \mathbf{B}_y, \mathbf{C}$ are the coefficient matrices, $k_x = -i\partial_x$, and $k_y = -i\partial_y$. For simplicity, we have considered only the terms up to the second order derivatives. With this Hamiltonian, the action integral corresponding to Eq.~(\ref{eq:kpsacteqn}) is given by
\begin{align}
\mathcal{A} =& \int d^2\rho\,\mathbb{F}^{\dagger}\cdot\mathbf{L}_{kp}\cdot\mathbb{F} \nonumber\\&+ \int d^2\rho\,S(\rho',\phi')\,\eta_{\rm scale}\,\mathbb{F}^{\dagger}\cdot\mathbb{A}\,\frac{\delta(\rho-\rho')}{\rho'},
\end{align}
where $\mathbf{L}_{kp}$ is the Lagrangian operator given by
\begin{align}
\mathbf{L}_{kp} = \Big[\cev{\partial}_x\mathbf{A}_{xx}\vec{\partial}_x+\cev{\partial}_y\mathbf{A}_{yy}\vec{\partial}_y+\frac{1}{2}\left(\cev{\partial}_x\mathbf{A}_{xy}\vec{\partial}_y+\cev{\partial}_y\mathbf{A}_{xy}\vec{\partial}_x\right)\nonumber\\+\frac{i}{2}\left(\cev{\partial}_x\mathbf{B}_{x}-\mathbf{B}_{x}\vec{\partial}_x\right)+\frac{i}{2}\left(\cev{\partial}_y\mathbf{B}_{y}-\mathbf{B}_{y}\vec{\partial}_y\right)+\mathbf{C}\Big],\nonumber
\end{align}
where $\cev{\partial}$ and $\vec{\partial}$ represent the derivative operator acting on the functions appearing to the left and right side, respectively. This will ensure the correct operator ordering at the material interface \cite{lrr_wavefunction}. 

Across a material interface, the envelope wavefunction, and the probability current has to be continuous. The continuity of the envelop function is ensured by using Hermite interpolation polynomials in our calculations, which will have both first and second derivative continuity \cite{KBR}. The probability current is evaluated using a gauge-variational approach that is commonly used in high energy/particle physics. Following Gell-Mann and Levy \cite{gellmanlevy}, we perform a transformation of the envelop function with respect to an arbitrary gauge function of the form, $\displaystyle\mathbb{F}\rightarrow\mathbb{F}\,e^{i\Lambda(\mathbf{\rho})}$. With this transformation, we obtain the conserved current 
\begin{align}
{J}_x &= \mathbb{F}^{\dagger}\cdot\frac{\delta\,\mathbf{L}_{kp}}{\delta\,\partial_x \Lambda}\cdot\mathbb{F}, \nonumber\\
{J}_y &= \mathbb{F}^{\dagger}\cdot\frac{\delta\,\mathbf{L}_{kp}}{\delta\,\partial_y \Lambda}\cdot\mathbb{F}. 
\end{align}
The interfacial potential is obtained from the DFT calculations, and will be continuous across the interface. Hence, this conserved current will also be continuous.

\section{Scale factor}\label{appdx:scale}
In this appendix, we obtain the expression for the scale factor $\eta_{\rm scale}$ employed in Eq.~(\ref{eq:kpsacteqn}). For simplicity, let us consider the $\textit{\textbf{k}}\cdot\textit{\textbf{p}}$ Hamiltonian in one-dimension (1D). The expression we obtain here for 1D translates into higher dimensions as well. Let us consider the equation
\begin{equation}
\left[\mathbf{H}_{kp}(x) -E\mathbf{1}\right]\cdot\mathbb{F} = S\,\eta_{\rm scale}\mathbb{A}\,\delta(x),
\end{equation}
where the source term in 1D is given by $S = 2ik_1$ \cite{sathwik1, ram_book}, with $k_1$ being the incoming wavevector. The solution to the above equation in terms of the Fourier components is given by
\begin{align}
\mathbb{F}(x) &= \frac{1}{2\pi}\int dk\,e^{ikx} 2ik_1\,\eta_{\rm scale}\,\frac{{\rm adj}(\mathbf{H}_{kp}-E\mathbf{1})\cdot\mathbb{A}}{{\rm Det}(\mathbf{H}_{kp}-E\mathbf{1})}.
\end{align}
Let us constrain $\mathbf{H}_{kp}$ to be a $2\times 2$ matrix, with eigenenergies given by $E_c$ and $E_v$. Then 
\begin{align}
\mathbb{F}(x)= \frac{ik_1}{\pi}\,\mathbb{A}\,{(E_c - E_v)}\int dk\,\frac{e^{ikx}\eta_{\rm scale}}{(k^2-k_1^2)(k^2-k_2^2)...(k^2-k_N^2)},
\end{align}
where $k_1, k_2 ... k_N$ are the solution wavevector for the determinantal equation ${\rm Det}(\mathbf{H}_{kp}-E\mathbf{1}) = 0$. The Hamiltonian $\mathbf{H}_{kp}$ can include terms in higher order $k$ (such as $k, k^2, k^3 ...$) and hence, the determinant equation will have $N$ roots. 
Therefore, we choose 
\begin{equation}\label{eq:scalingexp}
\eta_{\rm scale} = \frac{(k^2-k_2^2)...(k^2-k_N^2)}{(E_c-E_v)},
\end{equation}
so that the the incoming wavefunction $\psi(x)$ will have the form
\begin{equation}
\psi(x) = \mathbb{A}e^{ik_1 x}.
\end{equation} 
The expression derived in Eq.~(\ref{eq:scalingexp}) follows for 2D and higher dimensions as well.

\end{document}